\documentclass[12pt]{article}

\usepackage[margin=1in]{geometry}
\usepackage{amsfonts, amsmath, amssymb}
\usepackage[none]{hyphenat}
\usepackage{fancyhdr}
\usepackage{graphicx}
\usepackage{float}
\usepackage{graphicx}

\usepackage{array}

\usepackage{listings} 
\usepackage{longtable}
\usepackage{natbib}
\begin{document}

\begin{titlepage}
\begin{center}

\vspace*{4cm}
\line(1,0){450}\\
\Large{The R package \textbf{predint}}\\
\large{Prediction intervals for overdispersed binomial and
Poisson data, or based on linear random effects models in R}
\line(1,0){450}\\
\vspace*{2cm}
Max Menssen \\
Department of Biostatistics \\
Leibniz Universit\"{a}t Hannover \\
Herren\"{a}user Stra{\ss}e 2 \\
30419 Hannover \\

\vspace*{4cm}
\today

\end{center}
\end{titlepage}


\section*{Abstract}

A prediction interval is a statistical interval that should encompass one (or more)
future observation(s) with a given coverage probability and is usually computed based on 
historical control data. The application of prediction 
intervals is discussed in many fields of research, such as toxicology, pre-clinical
statistics, engineering, assay validation or for the assessment of replication studies.
Anyhow, the prediction intervals implemented in \textbf{predint} descent from previous work
that was done in the context of toxicology and pre-clinical applications. Hence
the implemented methodology reflects the data structures that are common in these 
fields of research.
In toxicology the historical data is often comprised of dichotomous or counted 
endpoints. Hence it seems natural to model these kind of data based on the binomial
or the Poisson distribution. Anyhow, the historical control data is usually comprised
of several studies. These clustering gives rise to possible overdispersion 
which has to be reflected for interval calculation.
In pre-clinical statistics, the endpoints are often assumed to be normal distributed, 
but usually are not independent from each other due to the experimental design
(cross-classified and/or hierarchical structures). These dependencies can be modeled
based on linear random effects models.
Hence, \textbf{predint} provides functions for the calculation of prediction intervals 
and one-sided bounds for overdispersed binomial data, for overdispersed Poisson 
data and for data that is modeled by linear random effects models.

\section*{Keywords}
Bio-assay, historical control data, bootstrap calibration, assay 
validation and qualification

\newpage



\section[Introduction]{Introduction} \label{sec:intro}

A prediction interval is a statistical interval $[L, U]$ that should encompass 
$M \geq 1$ future observations $\boldsymbol{y^*}$ simultaniously with coverage
probability 
\begin{equation*}
	P(L \leq \boldsymbol{y^*} \leq U) = 1 - \alpha. 
\end{equation*}
Similarly, lower prediction bounds $L$ should result in the coverage probability 
\begin{equation*}
	P(L \leq \boldsymbol{y^*} ) = 1 - \alpha 
\end{equation*}
and upper prediction bounds $U$ should have a coverage probability of
\begin{equation*}
	P(\boldsymbol{y^*} \leq U) = 1 - \alpha. 
\end{equation*}
The computation of prediction intervals or bounds is based on the assumption that
both, the historicalobservations $\boldsymbol{y}$ as well as the future observations 
$ \boldsymbol{y^*}$ descent from the same data generating process. \\
The application of the prediction intervals implemented in \textbf{predint}
is of use in several fields of research, such as the detection of anti-drug antibodies
\citep{Hoffmann+Berger:2011, Menssen+Schaarschmidt:2021} or
the validation of an actual control group by historical control data in toxicology 
\citep{Menssen+Schaarschmidt:2019}. Further applications of prediction intervals
can be found in industry \citep{Ryan:2007}, in experiments for method 
comparison (bridging) or assay validation \citep{Francq_et_al:2019} or in the context of 
the evaluation of replication studies \citep{Spence+Stanley:2016}.\\
All the applications mentioned above have in common, that the historical data used
for the calculation of prediction intervals is comprised of several clusters, rather
than of one unstructured sample. 
For example, in pre-clinical experiments for the detection of anti-drug antibodies, it is of 
interest to distinguish between 'responders' whose anti-drug antibody reaction 
exceeds a critical level and 'non-responders' whose reaction is uncritical low. 
One approach for the detection of such a critical level is the application of 
an upper prediction bound that is calculated based on the observed anti-drug 
antibody reaction of a set of known 'non-responders'. \\
Such experiments are usually run based on blood samples obtained from different 
patients (or animals) and might be analysed 
by different experimentors in different laboratories \citep{Hoffmann+Berger:2011}.
Since the experimental design is not of interest for answering the scientific question,
but reflects dependencies between experimental units (e.g. the samples that were 
analysed in the same laboratory are not independent from each other due to 
systematic error), random effects models have to be applied for modelling 
and interval calculation. \\
The prediction intervals implemented in \textbf{predint}, that are based on random 
effects models, are similar to the methodology proposed in \citet{Menssen+Schaarschmidt:2021}.
But, additionally to the historical experimental design, two of the three implementations are 
also able to take the design of the future data into account, rather than only the 
number of future observations as proposed in \citet{Menssen+Schaarschmidt:2021}.
Furthermore, the implemented methodology is applicable to a broad range of experimental
designs such as cross-classified and/or hierachical structures as well as balanced or 
unbalanced data (see section \ref{sec::ran_eff_mod}).\\
Another example where clustered data occurs are bio-assays with a toxicological
background. These experiments are usually comprised of an untreated control group 
that is compared to several groups of model organisms treated with a chemical 
compound of interest. 
In that field of research prediction intervals are of interest in order to validate 
the outcome of an actual (or future) control group. Hence, such intervals are 
calculated based on observations obtained from historical control groups of 
previous experiments \citep{Menssen+Schaarschmidt:2019, Valverde-Garcia:2018}.
Since many endpoints in toxicology are either dichotomous (e.g. rats with a tumor 
vs. rats without a tumor) or counted observations (e.g. numbers of eggs per hen), 
it seems natural to model them based on the binomial or the Poisson distribution,
respectively.
Anyhow, the model has to take the clustering into acount and hence also 
possible overdispersion, meaning that the variance of the data exceeds the variance
that can be modeled based on 'simple' binomial or Poisson distribution. One reason
for the presence of overdispersion are positive correleations between experimental units
in each cluster.  Therefore, overdispersion is considered to be almost always present 
in biological data \citep{Demetrio:2014, Mc_Cullagh+Nelder:1989}.\\
Prediction intervals for $M=1$ future observation based on overdispersed binomial data 
were proposed in \citet{Menssen+Schaarschmidt:2019}. Anyhow, the prediction intervals
for that type of data that are implemented in \textbf{predint} are based on a slightly 
different approach, meaning that they are also applicable in the case where
a simultainious prediction interval that should cover  $M>1$ future observations
is needed (see section \ref{sec::overdisp_bin_data}).\\
To the authors knowledge, prediction intervals for $M \geq 1$ future observations 
that can be computed based on clustered count data that exhibits ovsrdispersion were
not available in an \textbf{R} package hosted on CRAN. This gab is filled by 
the prediction interval proposed in section \ref{sec::overdisp_pois_data}.
%
%
%
%
\section{Theory} \label{sec:models}

\subsection{Linear random effects models} \label{sec::ran_eff_mod}

A general linear random effects model is given by
\begin{equation*}
\boldsymbol{Y} = \boldsymbol{1} \mu + \boldsymbol{Z} \boldsymbol{U} + \boldsymbol{e}
\end{equation*}
where $\boldsymbol{Y}= (Y_1, \ldots , Y_N)^T$ is a vector of random variables 
representing $n= 1, \ldots, N$ observations. The overall mean is represented by 
$\mu$. $\boldsymbol{U}$ is a stacked vector consisting of random effects sub-vectors 
$\boldsymbol{U}_c$. Hence, each $\boldsymbol{U}_c$ represents all $f=1, \ldots F_c$ 
levels associated with a particular random factor out of the $c= 1, \ldots, C$ random 
factors that influence the observations. $\boldsymbol{Z}$ is a design matrix of
dimensions $N \times F$ where $F= \sum_c F_c$ denotes the total length of
$\boldsymbol{U}$. The random errors associated with the observations are represented 
by $\boldsymbol{e}$. The individual random effects can be represented as $\boldsymbol{Z}_c \boldsymbol{U}_c$ such that 
\begin{equation*}
\boldsymbol{Z} \boldsymbol{U} = 
\begin{pmatrix}
	\boldsymbol{Z}_1 \cdots \boldsymbol{Z}_C
\end{pmatrix}
\begin{pmatrix}
	\boldsymbol{U}_1 \\
	\vdots \\
	 \boldsymbol{U}_C 
\end{pmatrix}
=
\sum_{c=1}^C \boldsymbol{Z}_c \boldsymbol{U}_c
\end{equation*}
with each
\begin{equation*}
\boldsymbol{U}_c =
\begin{pmatrix}
	U_{c1} \\
	\vdots \\
	U_{cF_c}
\end{pmatrix}
\end{equation*}
All random effects are assumed to be normal distributed $\boldsymbol{U}_c \sim N(\boldsymbol{0}, \boldsymbol{I} \sigma^2_c)$ as well as the errors $\boldsymbol{e} \sim N(\boldsymbol{0}, \boldsymbol{I} \sigma^2_{C+1})$. Furthermore it is assumed that
\begin{gather*}
        cov(\mu, U_{cF_c})=0 \quad \forall \quad c= 1, \ldots , C+1 \\
        cov(U_{cF_c}, U_{c'F_{c'}}) =0 \quad \forall \quad c= 1, \ldots , C+1, 
        \quad c' = 1, \ldots , C+1: c \neq c'
\end{gather*}
This model implies that the observations follow a multivariate normal distribution 
\begin{equation}
\boldsymbol{Y} \sim MVN(\boldsymbol{1} \mu, \boldsymbol{V}) \label{eq::Y_mvn}
\end{equation}
with variance-covariance matrix
\begin{equation*}
\boldsymbol{V} = \sum_{c=1}^C \boldsymbol{Z}_c \boldsymbol{Z}_c^T \sigma^2_c + \boldsymbol{I} \sigma^2_{C+1}.
\end{equation*}
It is assumed that both, the historical as well as the future random variables
are independent from each other, but descent from the same data generating process.
Hence, also the future random variable $\boldsymbol{Y^*}$ that represents 
$m=1, \ldots, M$ future observations is multivariate normal
\begin{equation}
\boldsymbol{Y^*} \sim MVN(\boldsymbol{1} \mu, \boldsymbol{V^*}) \label{eq::Y_star_mvn}
\end{equation}
with variance-covariance matrix
\begin{equation*}
\boldsymbol{V^*} = \sum_{c=1}^C \boldsymbol{Z}_c^{\boldsymbol{*}} \boldsymbol{Z}_c^{\boldsymbol{*}T} \sigma^2_c + \boldsymbol{I^*} \sigma^2_{C+1}.
\end{equation*}
Please note, that  the number of random effects per random factor might differ 
between the historical and the future data (e.g. 5 hospitals, each with 3 patients
vs. 3 hospitals, each with 4 patients). Consequently, the number of observations 
might differ between the historical and the future data ($\boldsymbol{Y}$ is of
length $N$ and $\boldsymbol{Y^*}$ is of length $M$) as well as the variance-covariance
matrices $\boldsymbol{V}$ and $\boldsymbol{V^*}$, since they depend on different
effects design matrices $\boldsymbol{Z}_c$ and $\boldsymbol{Z}^*_c$.\\
This implies that, $\boldsymbol{Y}$ and $\boldsymbol{Y^*}$ usually do not follow the
same multivariate normal distribution (see eq. \ref{eq::Y_mvn} and \ref{eq::Y_star_mvn}), 
but, nevertheless, descent from the same data generating process which depends 
only on the mean $\mu$ and the variance components $\sigma^2_c$. \\
In this setup, the error margin of the prediction is 
\begin{equation*}
\boldsymbol{D}=\boldsymbol{Y^*} - \boldsymbol{1} \mu
\end{equation*}
with 
\begin{gather*}
\boldsymbol{D} \sim MVN(\boldsymbol{0}, var(\boldsymbol{D}))\\
var(\boldsymbol{D}) = var(\boldsymbol{Y^*} - \boldsymbol{1} \mu) = var(\boldsymbol{Y^*})=\boldsymbol{V^*}
\end{gather*}
which, in the case of a prediction for only $M=1$ future observation simplifies to 
\begin{equation*}
\boldsymbol{V^*}=\sum_{c=1}^{C+1} \sigma^2_c
\end{equation*}
If a prediction interval for $M \geq 1$ future observations should be computed 
based on observed historical data $\boldsymbol{y}$ and the fitted model 
\begin{equation*}
\boldsymbol{y} = \boldsymbol{1} \hat{\mu} + \boldsymbol{Z} \hat{\boldsymbol{U}} + \hat{\boldsymbol{e}}
\end{equation*}
the estimated prediction variance becomes
\begin{equation*}
\widehat{var}(\boldsymbol{D}) = \widehat{var}(\boldsymbol{Y^*} - \boldsymbol{1} \hat{\mu}) =
\boldsymbol{V^*} +  \boldsymbol{J}\widehat{var}(\hat{\mu})
\end{equation*}
with $\boldsymbol{J}$ as a square matrix with all entries set to one.\\
A prediction interval that should cover $M > 1$ future observations $\boldsymbol{y^*}$
simultaniosly with coverage probability $1-\alpha$ is given by
\begin{equation}
\label{eqn:PIM>1}
[L, U] = \hat{\mu} \pm mvt_{1-\alpha/2, df, \widehat{var}(\boldsymbol{D})}
\end{equation}
with $mvt_{1-\alpha/2, df, \widehat{var}(\boldsymbol{D})}$ as the $1-\alpha/2$-quantile of the multivariate t-distribution. Please note, that a prediction interval for $M=1$ 
future observation simplifies to
\begin{equation}
\label{eqn:PIM1}
[L, U] = \hat{\mu} \pm t_{1-\alpha/2, df} \sqrt{\sum_{c=1}^{C+1} \hat{\sigma}^2_c + \widehat{var}(\hat{\mu})}.
\end{equation}
\citet{Menssen+Schaarschmidt:2021} gave an overview about several methods for the 
computation of the prediction intervals given in equations \ref{eqn:PIM>1} and \ref{eqn:PIM1}
of which their bootstrap calibrated prediction interval serves as the basis of the intervals 
implemented in \texttt{predint}.\\


\subsection{Overdispersed binomial data} \label{sec::overdisp_bin_data}

In several bio-assays run in the field of toxicology, the endpoints are dichotomous 
(e.g. rats with tumors vs. rats without tumors). A natural approach for modeling
such data is the binomial assumption
\begin{gather}
        y_h \sim bin(n_h, \pi) \label{eq.bin} \\ 
        var(y_h) = n_h \pi (1-\pi) \nonumber
\end{gather}
with $E(y_h)=n_h \pi$. In this notation $\pi$ is the binomial proportion, 
$n_h$ is the size of $h=1 \ldots H$ clusters (e.g. number of individuals in the 
$h$th historical control group) and $y_h$ are the number of successes obtained from the 
individuals of the $h$th cluster (e.g. rats with tumors). \\
Anyhow, most of the biological data that is assumed to be binomial has higher 
variability than expected and hence exhibits overdispersion \citep{Demetrio:2014, 
Mc_Cullagh+Nelder:1989}.\\
There are two approaches to model overdispersion: The quasi-binomial (or quasi-likelihood)
approach or modelling based on the beta-binomial distribution.
The first approach assumes a dispersion parameter that constantly inflates the variance 
for all observations, such that 
\begin{gather*}
	var(y_h)^{QB}= \phi n_h \pi (1-\pi)\label{eq::varQB}
\end{gather*}
with $E(y_h)=n_h \pi$ and $\phi > 1$.\\
For the latter, the data is assumed to be beta-binomial distributed
\begin{gather}
	\pi_h \sim beta(a, b) \nonumber \\
	y_h \sim bin(\pi_h, n_h) \nonumber \\
	var(y_h)^{BB}= n_h \pi (1-\pi) [1+(n_h-1) \rho] \label{eq::varBB}
\end{gather}
with $E(\pi_h)=\pi=a / (a+b)$, $E(y_h)=n_h \pi$ and $\rho=1/(1+a+b)$. It is noteworthy,
that $[1+(n_h-1) \rho]$ depends on the cluster size $n_h$ and becomes a constant, 
if all of the $H$ clusters have the same size $n_h=n_{h'}=n$. In this case the 
quasi-likelihood approach and the model that is based on the beta-binomial distribution
both result in overdispersion that constantly inflates the binomial variance
of the different clusters.\\
Several methods for the calculation of prediction intervals based on one binomial
sample were proposed in literature  \citep{Hahn+Meeker+Escobar:2017}. Anyhow,
none of these methods reflect the fact that the historical data is usually comprised of several 
clusters. Furthermore, these methods do not consider for possible overdispersion 
and hence, yield coverage probabilities far below the nominal level, if overdispersion
is present in the data \citep{Menssen+Schaarschmidt:2019}.\\
The prediction intervals for dichotomous data that are implemented in \textbf{predint}
are derived from an asymptotic prediction interval for $M=1$ future observation,
which is based on one unclustered binomial sample \citep{Hahn+Meeker+Escobar:2017}.
Its calculation is based on the assumption that
\begin{equation*}
	\frac{\hat{y}^* - Y^*}{\sqrt{\widehat{var}(\hat{y}^* - Y^*)}} = 
	\frac{n^* \hat{\pi} - Y^*}{\sqrt{\widehat{var}(n^* \hat{\pi} - Y^*)}} = 
	\frac{n^* \hat{\pi} - Y^*}{\sqrt{\widehat{var}(n^* \hat{\pi}) + \widehat{var}(Y^*)}} 
\end{equation*}
approximately follows a standard normal distribution
\begin{equation*}
        \frac{n^* \hat{\pi} - Y^*}{\sqrt{\widehat{var}(n^* \hat{\pi}) + \widehat{var}(Y^*)}} 
        \stackrel{appr.}{\sim} N(0,1).
\end{equation*}
In this notation $\hat{y}^*$ is the expected future observation, $Y^*$ is the future
random variable, $\hat{\pi}$ is the estimate for the binomial proportion obtained 
from the historical sample of size $n$ and $n^*$ is the size of the future cluster. 
The corresponding prediction interval for $M=1$ future observation is given by
\begin{equation}
	[l, u] = n^* \hat{\pi} \pm z_{1-\alpha/2} 
	\sqrt{n^* \hat{\pi} (1-\hat{\pi}) \Big(1 + \frac{n^*}{n} \Big)} 
	\label{eq::pi_bin}
\end{equation}
with $\widehat{var}(Y^*)=n^* \hat{\pi} (1-\hat{\pi})$ and $\widehat{var}(n^* \hat{\pi}) =
n^{*2} [\hat{\pi} (1-\hat{\pi}) /n]$.
As mentioned above, this interval was proposed for the application to one historical 
set of unclustered observations. Therefore, this prediction interval does not
account for clustering and hence, neglects the possible effect of overdispersion
that might occur in the data.\\
Prediction intervals for $M=1$ future observation, that account for the clustered 
structure of the historical data, can be calculated based on both, the quasi-binomial
approach or the beta-binomial distribution.
A prediction interval that is based on the quasi-binomial assumption 
can be obtained by substituting $\widehat{var}(Y)$ with $\widehat{var}(Y)^{QB} =\phi n^* \hat{\pi} (1-\hat{\pi})$ and $\widehat{var}(n^* \hat{\pi})$ with $\widehat{var}(n^* \hat{\pi})^{QB}=n^{*2} \phi \hat{\pi} (1-\hat{\pi}) / N$ in eq. \ref{eq::pi_bin}. Hence, the interval is defined as
\begin{equation}
	[l, u] = n^* \hat{\pi} \pm z_{1-\alpha/2} \sqrt{\phi  n^* \hat{\pi} (1-\hat{\pi}) \Big(1 + \frac{n^*}{N} \Big)}. \label{eq:quasi_bin}
\end{equation}
with $N=\sum_{h=1}^{H} n_h$. \\
A prediction interval for one future observation that is based on the beta-binomial distribution is computed if $\widehat{var}(Y)$ is substituted by $\widehat{var}(Y)^{BB}=n^* \hat{\pi} (1-\hat{\pi}) [1+(n^*-1) \hat{\rho}]$ and $\widehat{var}(n^* \hat{\pi})$ by $\widehat{var}(n^* \hat{\pi})^{BB}=\frac{n^{*2}\hat{\pi} (1-\hat{\pi})}{N} + \frac{N-1}{N} n^{*2} \hat{\pi} (1-\hat{\pi}) \hat{\rho}$ in eq. \ref{eq::pi_bin}. The resulting prediction interval is given as
\begin{equation}
	[l, u] = n^* \hat{\pi} \pm z_{1-\alpha/2} \sqrt{n^* \hat{\pi} (1-\hat{\pi}) \big[1+(n^*-1) \hat{\rho} \big] + \Big[ \frac{n^{*2}\hat{\pi} (1-\hat{\pi})}{N} + \frac{N-1}{N} n^{*2}\hat{\pi} (1-\hat{\pi}) \hat{\rho} \Big]} \label{eq::pi_beta_bin_asymp}
\end{equation}
Please note, that prediction intervals that should cover $M>$ future observations
simultainiously, can be obtained by the application of the bootstrap-calibration 
procedure described below in section \ref{sec::bs_calibration}. 


\subsection{Overdispersed Poisson data}  \label{sec::overdisp_pois_data}

In several bio-assays, such as avian reproduction, the variable of interest is 
comprised of count data \citep{Valverde-Garcia:2018}. A natural approach for
modeling counts is to assume them to be Poisson distributed
\begin{gather*}
	y_{h} \sim Pois(\lambda)\\
	E(y_{h}) = var(y_{h}) = \lambda.
\end{gather*}
Here, $y_{h}$ are the observations per cluster, $h=1 \ldots H$ is the index for 
the clusters and $\lambda$ is the Poisson mean. Similar to dichotomous data, 
overdispersion is usually present and can be modeled based on the quasi-Poisson
(quasi-likelihood) approach that grounds on a constant dispersion parameter 
inflating the Poisson-variance \citep{Demetrio:2014}, such that
\begin{gather*}
	var(y_{h})^{QP} = \phi \lambda
\end{gather*}
with $\phi > 1$ and $E(y_{h})=\lambda$.
Another approach for modeling overdispersed
Poisson data is the negative-binomial distribution where the means of the historical
studies follow a gamma distribution with parameters $a$ and $b$, such that 
\begin{gather*}
	\lambda_h \sim gamma(a, b)\\
	y_{h} \sim Pois(\lambda_{h})\\
	var(y_{h})^{NB} = \lambda + \kappa \lambda^2 = (1 + \kappa \lambda) \lambda
\end{gather*}
with $E(y_{h}) = \lambda = a/b$ and $\kappa=1/a$ \citep{Gsteiger:2013}. 
Please note that in the case in which several counted observations $y_{h}$ simply
vary around their expected value $\lambda$, both, the quasi-Poisson and the
negative-binomial assumption are not in contradiction with each other (with regard
to their variance formula). This is because both, $\phi$ and $(1 + \kappa \lambda)$ 
are constant in this case. Hence, 
\begin{equation}
	var(y_{h})^{NB} = var(y_{h})^{QB} = (1 + \kappa \lambda) \lambda  = \phi \lambda.
	\label{eq::var_QB_NB}
\end{equation}
Several methods for the calculation of prediction intervals for one future observation 
based on one Poisson distributed historical sample are reviewed in 
\citep{Hahn+Meeker+Escobar:2017}.
An asymptotic prediction interval for $M=1$ future observation $y^*$ which is based
on one unclustered Poisson distributed sample is based on the assumption that 
\begin{equation}
	\frac{\hat{y}^* - Y^*}{\sqrt{\widehat{var}(\hat{y}^* - Y^*)}} = 
	\frac{\hat{\lambda} - Y^*}{\sqrt{\widehat{var} \big({\hat{\lambda} - Y^* \big)}}} = 
	\frac{\hat{\lambda} - Y^*}{\sqrt{\widehat{var}({\hat{\lambda}) + \widehat{var}(Y^*)}}}  
	\label{eq::standard_pois}
\end{equation}
is approximately standard normal
\begin{equation*}
        \frac{\hat{\lambda} - Y^*}{\sqrt{\widehat{var}({\hat{\lambda}) + \widehat{var}(Y^*)}}} 
        \stackrel{appr.}{\sim} N(0,1).
\end{equation*}
The corresponding asymptotic prediction interval is given by
\begin{equation*}
	[l,u]= \hat{\lambda} \pm z_{1-\alpha/2}  \sqrt{2\hat{\lambda}}.
\end{equation*}
Please note that this interval is a simplified version (ignoring offsets) of the 
one that is reviewd in \citep{Hahn+Meeker+Escobar:2017}.
Its adaption to overdispersed data comprised of $h=1, \ldots , H$ clusters results in
\begin{equation}
	[l,u]=\hat{\lambda} \pm z_{1-\alpha/2} \sqrt{\hat{\phi} \hat{\lambda} \Big(1 + \frac{1}{H} \Big)} \label{eq::quasi_pois}
\end{equation}
with $\hat{\phi} > 1$. Simultanious prediction intervals for $M>1$ future observations can
be obtained by the application of the bootstrap calibration procedure described
in the next section.


\subsection{Bootstrap calibration} \label{sec::bs_calibration}

The bootstrap-calibration of statistical intervals dates back to the late 1980ies.
The original approach proposed by \citet{Loh:1987}  was aimed to find a better value 
for the $\alpha$ with which an interval is calculated in order to bring the coverage
probability of the calibrated interval as close as possible to the nominal $1-\alpha$.
This approach is reviewed in \citet{Efron+Tibshirani:1994} and is sometimes called 
alpha-calibration.\\
Contrary to alpha-calibration, the bootstrap-calibration procedure used for the 
calculation of the prediction intervals implemented in \textbf{predint}, is aimed to
find a coefficient $\delta$ that directly replaces the t- or z-quantiles in 
eq. \ref{eqn:PIM1}, \ref{eq:quasi_bin}, \ref{eq::pi_beta_bin_asymp} and 
\ref{eq::quasi_pois} resulting in prediction intervals for $M \geq 1$ future 
observations
\begin{equation}
\label{eq::bs_calib_pi}
[L, U] = \hat{y}^* \pm \delta \sqrt{\widehat{var}(\hat{y}^*) + \widehat{var}(Y^*)}
\end{equation}
for which the coverage probability is as close as possible to the nominal level
\begin{equation}
P(L \leq \boldsymbol{y^*} \leq U)=1-\alpha
\end{equation}
In this notation, $\boldsymbol{y^*}$ is the vector of the $m=1, \ldots, M$ future 
observations, $\hat{y}^*$ is the estimate for the expected future observation and $\sqrt{\widehat{var}(\hat{y}^*) + \widehat{var}(Y^*)}$ is the prediction error.\\
Please note, that all prediction intervals implemented in \textbf{predint} are of 
the form given in eq. \ref{eq::bs_calib_pi}. For interval calculation, the estimates 
that correspond to $\hat{y}^*$, $\widehat{var}(\hat{y}^*)$ and $\widehat{var}(Y^*)$, 
which off cause depend on the chosen model, are simply plugged in. Bootstrap
calibrated prediction intervals can be obtained depending on the following algorithm:
\begin{enumerate}
	\item Fit a random effects model to the historical data set $\boldsymbol{y}$ 
	      in order to obtain the estimates $\hat{y}^*$,
	      $\widehat{var}(\hat{y}^*)$ and $\widehat{var}(Y^*)$
	\item Draw $B$ parametric bootstrap samples $\boldsymbol{y}^*_b$ that 
	      follow the same experimental design as the future data 
	\item Additionally, draw $b=1, \ldots B$ further bootstrap samples 
	      $\boldsymbol{y}_b$ that follow the same experimental design as the 
	      historical data.
	\item Fit the initial model to $\boldsymbol{y}_b$ in order to obtain $\hat{y}^*_b$,
	      $\widehat{var}(\hat{y}^*_b)$ and $\widehat{var}(Y^*_b)$.
	\item Choose a start-value for $\delta$.
	\item Calculate prediction intervals based on the bootstrapped estimates 
	      as $$[L, U]_b = \hat{y}^*_b \pm \delta  \sqrt{\widehat{var}(\hat{y}^*_b)
	      + \widehat{var}(Y^*_b)}.$$
	\item Calculate the coverage probability for the prediction intervals that 
	      correspond to the particular $\delta$ as 
	      \begin{equation}
	             \hat{\Psi}_\delta = \frac{\sum_{b=1}^B I_\delta}{B} \label{eq::bs_cp}
	      \end{equation}
	      with $I_\delta=1$ if $\boldsymbol{y}^*_b \in [L, U]_b$ and $I_\delta=0$ 
	      if $\boldsymbol{y}^*_b \notin [L, U]_b$.
	\item Alternate $\delta$ and repeat step six and seven until $\hat{\Psi}_\delta$ 
	      is satisfactory close to the nominal $1-\alpha$. Use this particular 
	      value of $\delta$ for the calculation of the calibrated prediction interval.
	\item Calculate the calibrated prediction interval based on the chosen 
	      $\delta$ and the parameter estimates from the initial model as 
	      shown in equation \ref{eq::bs_calib_pi}.
\end{enumerate}
Please note, that for all prediction intervals implemented in \textbf{predint} the 
search for $\delta$ in step 8 of the calibration algorithm depends on the following
bisection:
\begin{enumerate}
        \item Define start values $\delta_{min}$ and $\delta_{max}$ such that the
                corresponding bootstrap coverage probabilities $\hat{\Psi}_{\delta_{min}}$
                and $\hat{\Psi}_{\delta_{max}}$ estimated following eq. \ref{eq::bs_cp} 
                are
                \begin{gather*}
                        \hat{\Psi}_{\delta_{min}} < 1-\alpha \\
                        \hat{\Psi}_{\delta_{max}} > 1-\alpha.
                \end{gather*}
        \item Start the first of $g=1, \ldots, G$ bisection steps by defining
                \begin{equation*}
                        \delta_1 = \frac{\delta_{min} + \delta_{max}}{2}
                \end{equation*}
        \item Calculate the corresponding bootstrap coverage probability $\hat{\Psi}_{\delta_{1}}$
                according to eq. \ref{eq::bs_cp}
        \item If $\hat{\Psi}_{\delta_{1}} < 1-\alpha $, calculate
                \begin{equation*}
                        \delta_2 = \frac{\delta_{1} + \delta_{max}}{2}
                \end{equation*}
                If $\hat{\Psi}_{\delta_{1}} > 1-\alpha $, calculate
                \begin{equation*}
                        \delta_2 = \frac{\delta_{1} + \delta_{min}}{2}
                \end{equation*}
        \item Calculate the bootstrap coverage probability $\hat{\Psi}_{\delta_{2}}$
                according to eq. \ref{eq::bs_cp}
        \item Repeat this iteration process until $\mid 1-\alpha - \hat{\Psi}_{\delta_{G}} \mid < t$
                or a maximum number of bisection steps $G_{max}$ was done.
        \item Use this particular $\delta_{G}$ for the calculation 
                of the calibrated interval (repalace $\delta$ by $\delta_{G}$ in 
                eq. \ref{eq::bs_calib_pi}).
\end{enumerate}


\section{Methodology implemented in predint}

\subsection{Prediction intervals}
Since all functions for the calculation of prediction intervals (see tab. \ref{table:PI})
depend on the same calibration approach and the intervals are of the same form 
(see eq. \ref{eq::bs_calib_pi}), all functions share a common framework in terms
of applicability. The arguments common to all functions for interval calculation 
are given in tab. \ref{table:arguments}.
\begin{table}[ht]
\centering
\caption{Functions for the calculation of prediction intervals}
\label{table:PI}
\begin{tabular}{l|l}
\hline
 \textbf{Function name} & \textbf{Functionality}  \\ \hline
 \texttt{lmer\_pi\_unstruc()} &    \\
 \texttt{lmer\_pi\_futvec()} & PI based on random effects modeles  \\
 \texttt{lmer\_pi\_futmat()} &    \\ \hline
 \texttt{beta\_bin\_pi()} &  PI for overdispersed binomial data  \\
 \texttt{quasi\_bin\_pi()} &    \\ \hline
 \texttt{quasi\_pois\_pi()} &  PI for overdispersed count data   \\ \hline
\end{tabular}
\end{table}

\begin{figure}[ht]\centering
	\includegraphics[width=\textwidth, clip]{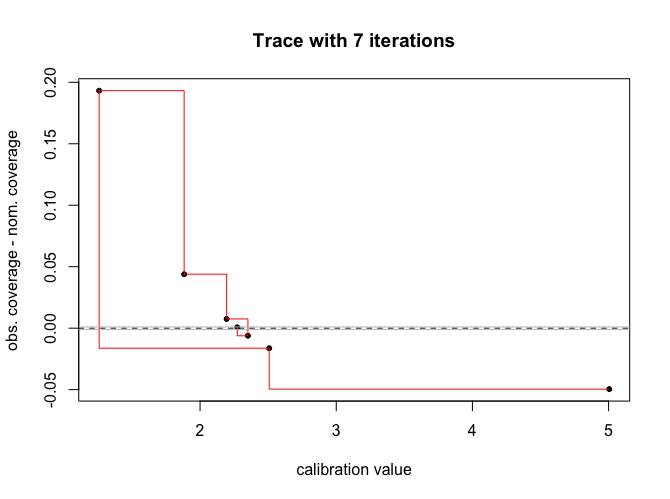}
	\caption{Grafical overview about the bisection steps}
	\label{fig::traceplot}
\end{figure}

\begin{table}[ht]
\centering
\caption{Arguments common to all functions for interval calculation}
\label{table:arguments}
\begin{tabular}{l|l}
\hline
 \textbf{Argument} & \textbf{Functionality}  \\ \hline
 \texttt{alternative} &  Prediction intervals or bounds $L$, $U$ \\ \hline
 \texttt{alpha} &  Definition of $\alpha$  \\ \hline
 \texttt{nboot} &  Number of bootstrap samples $B$ \\ \hline
 \texttt{delta\_min} &  Lower start value for bisection $\delta\_{min}$ \\ \hline
 \texttt{delta\_max} &  Upper start value for bisection $\delta\_{max}$ \\ \hline
 \texttt{tolerance} &  Tolerance for bisection $t$ \\ \hline
 \texttt{traceplot} &  Graphical overview about the bisection \\ \hline
 \texttt{n\_bisec} &  Max. number of bisection steps $G\_{max}$ \\ \hline
\end{tabular}
\end{table}
Prediction intervals are calculated with \texttt{alternative="both"}, which is the
default setting. Anyhow, if lower prediction bounds are of interest, \texttt{alternative}
has to be set to \texttt{"lower"}. Upper prediction bounds are computed with
\texttt{alternative="upper"}. Please note, that bootstrap calibration of prediction
bounds is done by adopting eq. \ref{eq::bs_cp}, where $L=-\infty$, if \texttt{alternative}
is set to \texttt{"upper"}, or $U=\infty$ if \texttt{alternative} is set to \texttt{"lower"}.\\
If not specified explicitely, all functions calculate prediction intervals (or bounds)
with coverage probability $1-\alpha=0.95$ which can be alterated by setting \texttt{alpha}
to any value between 0 and 1.
The number of bootstrap samples can be specified by \texttt{nboot} wich is set to
10000 by default.
Start values for the bisection are provided by the arguments \texttt{delta\_min}
and \texttt{delta\_max} which are set to default values of 0.01 and 10. The maximum
number of bisection steps is controlled via \texttt{n\_bisec} and is per default 30. \\
A graphical overview about the bisection process is given if \texttt{traceplot=TRUE}
(see fig. \ref{fig::traceplot}). In such a plot, the calibration values $\delta_g$,
calculated during  the bisection, are given on the x-axis. The y-axix shows the
difference between the observed bootstrap coverage probabilities $\hat{\Psi}_{\delta_{g}}$
and the nominal level $1-\alpha$. The bisection stops if $1-\alpha - \hat{\Psi}_{\delta_{g}}
\in 0 \pm t$ or, if this is not the case, after the maximum number of steps defined
via \texttt{n\_bisec}.
In rare occasions it might happen, that the estimated coverage probabilities
$\hat{\Psi}_{\delta_g}$ do not converge to the nominal level $1-\alpha$. This happens
if $\hat{\Psi}_{\delta_{Gmax}} \notin [(1-\alpha)-t, (1-\alpha)+t]$ with $t$ controlled via
\texttt{tolerance}. In this case the value for $\delta_{Gmax}$ from the last bisection step
is chosen for interval calculation in eq. \ref{eq::bs_calib_pi}. The user can
decide either to use the calculated interval or to change the search-interval
for $\delta$ by changing \texttt{lambda\_min} and
\texttt{lambda\_max}. Alternatively one might increase the tolerable level around the
nominal coverage probability $(1-\alpha)$ via \texttt{tolerance}.\\
Please note, that due to the discretenes of dichotomous or count data, the true
coverage probability of the interval might not approach the desired $1-\alpha$ in
some occasions. Consequently, also the bisection might not converge to the
nominal level. In such cases the calibrated prediction interval corresponding to
the last of the $G_{max}$ bisection steps, should be the one with coverage
probability closest to the nominal level.


\subsubsection{Prediction intervals based on linear random effects models} \label{sec::PI_ranef}

Prediction intervals, that are based on random effects modeles fit with \texttt{lme4::lmer()}
to the historical data, can be computed using \texttt{lmer\_pi\_unstruc()}, \texttt{lmer\_pi\_futvec()}
or \texttt{lmer\_pi\_futmat()}.
These intervals depend on the historical mean $\hat{\mu}$ which is extracted
from the fitted model with \texttt{lme4::fixef()}, its estimated variance
$\widehat{var}(\hat{\mu})$ drawn from the fitted model with \texttt{lme4::vcov.merMod()}
and $\widehat{var}(Y^*)$, the sum of the variance components extracted from the
fitted model with \texttt{lme4::VarCorr()}. Substituting these estimates into eq.
\ref{eq::bs_calib_pi} results in a bootstrap calibrated prediction interval
\begin{equation}
[L, U] = \hat{\mu} \pm \delta \sqrt{\widehat{var}(\hat{\mu}) + \sum_{c=1}^{C+1} \hat{\sigma}^2_c}.
\end{equation}
This interval can be applied either in the case where a prediction for one future observation is needed,
as well as in the case where $M>1$ future observations should be predicted.\\
In the examples below, \texttt{c2\_dat1} will serve as an example for a
historical data set. It descents from a two way completely cross-classified design
with three replications per random factor and three replications per interaction
term and is therefore comprised of 27 observations.

\begin{verbatim}
R> c2_dat1

       y_ijk a b
1  105.27359 1 1
2  101.40640 1 1
3   94.01300 1 1
4   97.82988 2 1
5   94.30743 2 1
6   92.52234 2 1
7  102.17317 3 1
8   99.74908 3 1
9  100.64042 3 1
10  95.49433 1 2
11  92.30937 1 2
12  99.88281 1 2
13 103.82970 2 2
14  99.95517 2 2
15 107.13102 2 2
16 107.42282 3 2
17 105.25822 3 2
18 108.82881 3 2
19 107.30048 1 3
20 107.13083 1 3
21 106.73200 1 3
22 106.44846 2 3
23 104.60098 2 3
24 103.86882 2 3
25 107.01238 3 3
26 106.06968 3 3
27 107.53004 3 3

\end{verbatim}
A random effects model that reflects the experimental design of \texttt{c2\_dat1},
can be fitted with
\begin{verbatim}
R> # install.packages("lme4")
R> library(lme4)
R> fit <- lmer(y_ijk~(1|a)+(1|b)+(1|a:b), data=c2_dat1)
\end{verbatim}
In all three functions, the fitted model has to be specified via \texttt{model}.
Please note, that at the current state, only models
in which the random effects are specified as \texttt{(1|random effect)} are
supported.\\
The bootstrap sampling of future observations $\boldsymbol{y}^*_b$ is the same
in all three functions, if a prediction interval for $M=1$ future observation is
needed. This is because, internally, the future data is bootstrapped from the fitted model
via \texttt{lme4::bootMer()} of which one observation per bootstrap data set
is randomly chosen to serve as
$\boldsymbol{y}^*_b$ in step 2 of the calibration process. Hence all three functions
yield the same prediction interval in this case.
\begin{verbatim}
R> set.seed(1234)
R> lmer_pi_unstruc(model=fit, m=1, alternative="both", nboot=10000)
\end{verbatim}
\begin{verbatim}
  m hist_mean quant_calib pred_se  lower    upper
1 1  102.3971    2.273359 5.923724 88.93033 115.8638 
\end{verbatim} 

\newpage

\begin{verbatim}
R> set.seed(1234)
R> lmer_pi_futvec(model=fit, futvec=1, alternative="both", nboot=10000)
\end{verbatim}
\begin{verbatim}
  m hist_mean quant_calib  pred_se    lower    upper
1 1  102.3971    2.273359 5.923724 88.93033 115.8638
\end{verbatim}
\begin{verbatim}
R> set.seed(1234)
R> lmer_pi_futmat(model=fit, newdat=1, alternative="both", nboot=10000)
\end{verbatim}
\begin{verbatim}
  m hist_mean quant_calib  pred_se    lower    upper
1 1  102.3971    2.273359 5.923724 88.93033 115.8638
\end{verbatim}
The output of the three functions is a \texttt{data.frame} where \texttt{m} is the
number of future observations the prediction interval should cover (in this case one).
The historical mean $\hat{\mu}$ is given by \texttt{hist\_mean} and
\texttt{quant\_calib} is the bootstrap calibrated coefficient used for the calculation
of the interval ($\delta$ in eq. \ref{eq::bs_calib_pi}). \texttt{pred\_se} is the
estimated standard error of the prediction ($\sqrt{\widehat{var}(\hat{\mu}) + \widehat{var}( Y^*)}$
in eq. \ref{eq::bs_calib_pi}) and \texttt{lower} and \texttt{upper} are the lower and
the upper bounds of the prediction interval.\\
The only difference between the three functions is the way how the bootstrap samples
$\boldsymbol{y}^*_b$ are drwan, if a prediction interval for $M>1$ future observations
is needed. In the examples below, \texttt{predint::c2\_dat3} will serve as a future
data set that descents from the same data generating process, but has only two observations
per random factor and hence eight observations in total.\\
\begin{verbatim}
R> c2_dat3
      y_ijk a b
1  97.47232 1 1
2  95.44895 1 1
3 100.18817 2 1
4  99.36843 2 1
5  99.08363 1 2
6 101.11561 1 2
7  97.05361 2 2
8  97.81136 2 2
\end{verbatim}
%
\texttt{lmer\_pi\_unstruc()} is a direct implementation of the prediction interval
described in \citet{Menssen+Schaarschmidt:2021}. Hence, if $M>1$ the bootstrapped future
observations  $\boldsymbol{y}^*_b$ are sampled in two steps. Firstly, bootstrap samples
that have the same
experimental structure as the historical data are sampled using \texttt{lme4::bootMer()}.
Then, $M$ observations are drawn randomly from the bootstrapped data in order to serve
as $\boldsymbol{y}^*_b$ in the calibration. Therefore, only the number of future
observations, but not the experimental design of the future data set is
considered.\\
A prediction interval for $M=8$ future observations can be obtained, if \texttt{m}
is set to 8 or if the future data set \texttt{c2\_dat3} is directly specified
via \texttt{newdat}.
\begin{verbatim}
R> set.seed(1234)
R> lmer_pi_unstruc(model=fit, m=8, alternative="both", nboot=10000)
  m hist_mean quant_calib  pred_se    lower    upper
1 8  102.3971    3.366016 5.923724 82.45774 122.3364
\end{verbatim}
\begin{verbatim}
R> set.seed(1234)
R> lmer_pi_unstruc(model=fit, 
                   newdat=c2_dat3, 
                   alternative="both",
                   nboot=10000)
      y_ijk a b hist_mean quant_calib  pred_se    lower    upper cover
1  97.47232 1 1  102.3971    3.366016 5.923724 82.45774 122.3364  TRUE
2  95.44895 1 1  102.3971    3.366016 5.923724 82.45774 122.3364  TRUE
3 100.18817 2 1  102.3971    3.366016 5.923724 82.45774 122.3364  TRUE
4  99.36843 2 1  102.3971    3.366016 5.923724 82.45774 122.3364  TRUE
5  99.08363 1 2  102.3971    3.366016 5.923724 82.45774 122.3364  TRUE
6 101.11561 1 2  102.3971    3.366016 5.923724 82.45774 122.3364  TRUE
7  97.05361 2 2  102.3971    3.366016 5.923724 82.45774 122.3364  TRUE
8  97.81136 2 2  102.3971    3.366016 5.923724 82.45774 122.3364  TRUE
\end{verbatim}
If \texttt{newdat} is specified, the output is a \texttt{data.frame} in which the
first columns represent the data set specified via \texttt{newdat}.
\texttt{hist\_mean, quant\_calib, pred\_se, lower} and \texttt{upper} are the same as
above. \texttt{cover} gives a statement whether the observation is covered
by the interval or not.\\
%
%
Contrary to \texttt{lmer\_pi\_unstruc()}, \texttt{lmer\_pi\_futvec()} accounts
for the experimental design of the future data and is applicable if the experimental
design of the future data is part of the design of the historical experiment(s).
If a prediction interval for
$M>1$ future observations is needed, a vector of row numbers that define the
experimental structure of the future data based on the historical data set has to be
specified.
\begin{verbatim}
R> futvec <- c(1, 2, 4, 5, 10, 11, 13, 14)
\end{verbatim}
defines the rows in \texttt{c2\_dat1} that correspond to the experimental design
of \texttt{c2\_dat3} (two observations per random factors a and b and their interaction).
In other words, if the observations defined by the row numbers
given in \texttt{futvec} are subsetted from \texttt{c2\_dat1}, these subset will
appear to descent from the same experimental design as \texttt{c2\_dat3}.
\begin{verbatim}
R> c2_dat1[futvec, ]
       y_ijk a b
1  105.27359 1 1
2  101.40640 1 1
4   97.82988 2 1
5   94.30743 2 1
10  95.49433 1 2
11  92.30937 1 2
13 103.82970 2 2
14  99.95517 2 2
\end{verbatim}
Internally, the bootstrap samples corresponding to the future
observations are sampled with \texttt{lme4::bootMer()}. Then for each of the
bootstrap samples, a subset that is comprised of the observations in the rows defined by
\texttt{futvec} is drawn and serves as $\boldsymbol{y}^*_b$ in the calibration.
A prediction interval for the 8 future observations
in \texttt{c2\_dat3} can be obtained with
\begin{verbatim}
R> set.seed(1234)
R> lmer_pi_futvec(model=fit, futvec=futvec, alternative="both", nboot=10000)
  m hist_mean quant_calib  pred_se    lower    upper
1 8  102.3971     3.30748 5.923724 82.80448 121.9897
\end{verbatim}
If the future data should appear in the output, it can be specified via \texttt{newdat}
but, of cause its data structure has to correspond to the structure defined by
\texttt{futvec}.
\begin{verbatim}
R> set.seed(1234)
R> lmer_pi_futvec(model=fit,
               futvec=futvec,
               newdat=c2_dat3,
               alternative="both",
               nboot=10000)
      y_ijk a b hist_mean quant_calib  pred_se    lower    upper cover
1  97.47232 1 1  102.3971    3.287969 5.923724 82.92006 121.8741  TRUE
2  95.44895 1 1  102.3971    3.287969 5.923724 82.92006 121.8741  TRUE
3 100.18817 2 1  102.3971    3.287969 5.923724 82.92006 121.8741  TRUE
4  99.36843 2 1  102.3971    3.287969 5.923724 82.92006 121.8741  TRUE
5  99.08363 1 2  102.3971    3.287969 5.923724 82.92006 121.8741  TRUE
6 101.11561 1 2  102.3971    3.287969 5.923724 82.92006 121.8741  TRUE
7  97.05361 2 2  102.3971    3.287969 5.923724 82.92006 121.8741  TRUE
8  97.81136 2 2  102.3971    3.287969 5.923724 82.92006 121.8741  TRUE
\end{verbatim}
Despite its user friendliness \texttt{lmer\_pi\_futvec()} has one drawback: It is only
applicable if the experimental design of the future data is included in the experimental
design of the historical data. In other words, \texttt{lmer\_pi\_futvec()} is not
applicable if the number of observations per random factor is bigger in the future
data than in the historical data (e.g. if data will be observed from four future
laboratories but the historical data contains only three historical ones). \\
This problem can be overcome by using \texttt{lmer\_pi\_futmat()}, which is the only
function in which the $M>1$ future observations are directly bootstraped from
the experimental design of the future data.
As already stated in section \ref{sec::ran_eff_mod}, a prediction intervall
for $M > 1$ future observations depends on the numbers of observations per
random factor in the future data set. Hence, the variance-covariance matrices
$\boldsymbol{V}$ and $\boldsymbol{V^*}$ for the historical and the future observations
differ from each other if $N \neq M$. \\
The bootstrap sampling used in \texttt{lmer\_pi\_futmat()} is based on the following
algorithm which is implemented in \texttt{lmer\_bs()} (see section \ref{sec::sampling}):
\begin{enumerate}
        \item{Obtain the estimates for the mean $\hat{\mu}$ and the variance
                components $\hat{\sigma}^2_{c+1}$ based on the model fit to the
                historical data set.}
        \item{Define the design matrices $\boldsymbol{Z}^*_c$ each of dimensions
                $M \times F^*_c$, with $M$ as the number of future observations and
                $F^*_c$ as the number of observations per random factor in the
                future data set.}
        \item{Draw random samples that correspond to the random effects, such that
                $\boldsymbol{U}^*_c \sim N(\boldsymbol{0}, \boldsymbol{I} \hat{\sigma}^2_c)$
                and $\boldsymbol{e}^* \sim N(\boldsymbol{0}, \boldsymbol{I} \hat{\sigma}^2_{C+1})$.}
        \item{Calculate the bootstrap sample as $\boldsymbol{y} = \boldsymbol{1} \hat{\mu} +
                \sum_{c=1}^{C} \boldsymbol{Z}^*_c \boldsymbol{U}^*_c + \boldsymbol{e}^*$.}
        \item{In order to obtain $B$ bootstrap samples, repeat step 1 to 4 for $b=1, \ldots, B$
                times.}
\end{enumerate}
If the future data is handed over via \texttt{newdat}, the bootstrap
depends on a list containing the design matrices $\boldsymbol{Z}^*_c$ that was
created using \texttt{lme4::lFormula()}. Hence each random factor in \texttt{newdat}
needs at least two replications. A prediction
interval for \texttt{c2\_dat3} is given with
\begin{verbatim}
R> set.seed(1234)
R> lmer_pi_futmat(model=fit,
               newdat=c2_dat3,
               alternative="both",
               nboot=10000)
      y_ijk a b hist_mean quant_calib  pred_se   lower    upper cover
1  97.47232 1 1  102.3971    3.326992 5.923724 82.6889 122.1053  TRUE
2  95.44895 1 1  102.3971    3.326992 5.923724 82.6889 122.1053  TRUE
3 100.18817 2 1  102.3971    3.326992 5.923724 82.6889 122.1053  TRUE
4  99.36843 2 1  102.3971    3.326992 5.923724 82.6889 122.1053  TRUE
5  99.08363 1 2  102.3971    3.326992 5.923724 82.6889 122.1053  TRUE
6 101.11561 1 2  102.3971    3.326992 5.923724 82.6889 122.1053  TRUE
7  97.05361 2 2  102.3971    3.326992 5.923724 82.6889 122.1053  TRUE
8  97.81136 2 2  102.3971    3.326992 5.923724 82.6889 122.1053  TRUE
\end{verbatim}
Sometimes a random factor in the future data set might not have any replicate e.g.
if the historical data descents from trials that were done in several different
laboratories, but the experiments for the future observations were carried out in
another one. This is the case in \texttt{c2\_dat4}, where the factor \texttt{b} has only
one observation.
\begin{verbatim}
R> c2_dat4
     y_ijk a b
1 102.8583 1 1
2 101.1324 1 1
3 104.9425 2 1
4 101.2299 2 1
5 104.6727 2 1
6 105.3402 2 1
\end{verbatim}
Here, the future data can not be specified via \texttt{newdat} since \texttt{lme4::lFormula()}
can not handle such cases. Alternatively, a list that contains the design matrices
$\boldsymbol{Z}^*_c$ can be provided via \texttt{futmat\_list}. Please note, that
the order of the design matrices has to correspond to the order by which the random
factors are handled in the initial model that was fit to the historical data
with \texttt{lme4::lmer()}.
A list of design matrices corresponding to \texttt{c2\_dat4} is given by
\begin{verbatim}
R> fml <- vector(length=4, "list")
R>
R> names(fml) <- c("a:b", "b", "a", "Residual")
R>
R> fml[["a:b"]] <- matrix(nrow=6, ncol=2,
                       data=c(1,1,0,0,0,0,
                              0,0,1,1,1,1))
R>
R> fml[["b"]] <- matrix(nrow=6, ncol=1,
                     data=c(1,1,1,1,1,1))
R>
R> fml[["a"]] <- matrix(nrow=6, ncol=2,
                     data=c(1,1,0,0,0,0,
                            0,0,1,1,1,1))
R>
R> fml[["Residual"]] <- diag(6)
R>
R> fml
\end{verbatim}
\begin{verbatim}
$`a:b`
     [,1] [,2]
[1,]    1    0
[2,]    1    0
[3,]    0    1
[4,]    0    1
[5,]    0    1
[6,]    0    1


$b
     [,1]
[1,]    1
[2,]    1
[3,]    1
[4,]    1
[5,]    1
[6,]    1

$a
     [,1] [,2]
[1,]    1    0
[2,]    1    0
[3,]    0    1
[4,]    0    1
[5,]    0    1
[6,]    0    1

$Residual
     [,1] [,2] [,3] [,4] [,5] [,6]
[1,]    1    0    0    0    0    0
[2,]    0    1    0    0    0    0
[3,]    0    0    1    0    0    0
[4,]    0    0    0    1    0    0
[5,]    0    0    0    0    1    0
[6,]    0    0    0    0    0    1

\end{verbatim}
The corresponding prediction interval is given by
\begin{verbatim}
R> set.seed(1234)
R> lmer_pi_futmat(model=fit, 
                  futmat_list=fml, 
                  alternative="both",
                  nboot=10000)
\end{verbatim}
\begin{verbatim}
  m hist_mean quant_calib  pred_se    lower    upper
1 6  102.3971    3.034316 5.923724 84.42263 120.3715
\end{verbatim}


\subsubsection{Prediction intervals for overdispersed binomial data}

Prediction intervals for overdispersed binomial data can be calculated based on
the quasi-likelihood aproach using \texttt{quasi\_bin\_pi()} or based on the
beta-binomial assumption using \texttt{beta\_bin\_pi()}.
Because overdispersion appeals as a constant if the cluster size is the same, \texttt{qb\_dat1}
will serve as an example for the historical data on which prediction intervals
will be calculated based on both assumptions. The data set is comprised of the
numbers of success (e.g. rats with tumors) vs. the number of failures (e.g. rats
without a tumor) obtained in 10 clusters, each comprised of 50 experimental untits
(e.g. rats).
\begin{verbatim}
R> qb_dat1
\end{verbatim}
\begin{verbatim}
   succ fail
1     0   50
2     9   41
3    13   37
4     1   49
5     4   46
6     5   45
7    13   37
8     7   43
9     7   43
10    6   44
\end{verbatim}
Based on the quasi-likelihood approach, \texttt{quasi\_bin\_pi()} calculates bootstrap
calibrated prediction intervals for $M \geq 1$ future numbers of success $y^*_m$
\begin{equation*}
	[l, u]_m = n_m^* \hat{\pi} \pm \delta \sqrt{\phi  n_m^* \hat{\pi} (1-\hat{\pi}) \Big(1 + \frac{n_m^*}{\sum_{h=1}^{H} n_h} \Big)}.
\end{equation*}
with $n_m^*$ as the size of $m=1, \ldots, M$ future clusters. \\
Please note, that the calculation of predciction intervals depend on the future
cluster size $n_m^*$ and hence, the calculated prediction intervals are different,
if the size of the future clusters differs beween each other.\\
The historical data set has to be specified \texttt{histdat} and needs to be a
\texttt{data.frame} with two columns, of which one describes the numbers of success
and the other the numbers of failures.
Then, internally, the estimation of $\hat{\phi}$ and $\hat{\pi}$ is done based on a
generalized linear model, fit with
\texttt{glm(cbind(histdat[,1],histdat[,2]) \~{} 1, family=quasibinomial(), data=histdat)}.
The bootstrap data used in step two and three of the calibration process described
above, is sampled using the \texttt{rqbinom()} function which is described in detail
in section \ref{sec::sampling}.\\
A prediction interval for the number of success in one future cluster of size 50
can be obtained with
\begin{verbatim}
R> set.seed(1234)
R> quasi_bin_pi(histdat=qb_dat1, newsize=50, nboot=10000)
\end{verbatim}
\begin{verbatim}
  total hist_prob quant_calib  pred_se lower    upper
1    50      0.13   0.9855859 10.72381     0 17.06923
\end{verbatim}
The resulting output is a \texttt{data.frame} in which \texttt{total} indicates the future
cluster size $n^*$, \texttt{hist\_prob} is the estimate for the historical binomial proportion
$\hat{\pi}$, \texttt{quant\_calib} is the bootstrap calibrated coefficient used for
the calculation of the interval ($\delta$ in eq. \ref{eq::bs_calib_pi}).
\texttt{pred\_se} is the prediction error $\sqrt{\phi n_m^* \hat{\pi} (1-\hat{\pi}) \Big(1 + n_m^* / \sum_{h=1}^{H} n_h \Big)}$
and the prediction interval is given by \texttt{lower} and \texttt{upper}.\\
Prediction intervals that simultainiously cover $M=3$ future numbers of success
which are observed in clusters of size 40, 50 and 60 can be calculated with
\begin{verbatim}
R> set.seed(1234)
R> quasi_bin_pi(histdat=qb_dat1, newsize=c(40, 50, 60), nboot=10000)
\end{verbatim}
\begin{verbatim}
  total hist_prob quant_calib  pred_se lower    upper
1    40      0.13    1.434355  8.75595     0 17.75915
2    50      0.13    1.434355 10.72381     0 21.88175
3    60      0.13    1.434355 12.68858     0 25.99993
\end{verbatim}
If the future data should appear in the output, it can be specified via \texttt{newdat}.
Please note, that the future data has to be of the same structure as the historical
one (two variables, one for success and one for failures). Defining \texttt{qb\_dat2}
via \texttt{newdat} results in the following output.
\begin{verbatim}
R> qb_dat2
\end{verbatim}
\begin{verbatim}
  succ fail
1    0   40
2    6   44
3    8   52
\end{verbatim}
\begin{verbatim}
R> set.seed(1234)
R> quasi_bin_pi(histdat=qb_dat1, newdat=qb_dat2, nboot=10000)
\end{verbatim}
\begin{verbatim}
  succ fail total hist_prob quant_calib  pred_se lower    upper cover
1    0   40    40      0.13    1.434355  8.75595     0 17.75915  TRUE
2    6   44    50      0.13    1.434355 10.72381     0 21.88175  TRUE
3    8   52    60      0.13    1.434355 12.68858     0 25.99993  TRUE
\end{verbatim}
In this output, three further variables occur: The first two variables are the
data set specified via \texttt{newdat}. \texttt{total} is the clustersite $n^*_m$
and \texttt{cover} gives a statement, if the prediction intervals cover their
corresponding future observation (first column of the output).\\
Bootstrap calibrated prediction intervals that are based on the beta-binomial
assumption, can be computed with \texttt{beta\_bin\_pi()}. The resulting prediction
intervals are given as
\begin{equation*}
	[l, u]_m = n_m^* \hat{\pi} \pm \delta \sqrt{n_m^* \hat{\pi} (1-\hat{\pi})
	\big[1+(n_m^*-1) \hat{\rho} \big] + \Big[ \frac{n_m^{*2}\hat{\pi} (1-\hat{\pi})}{N} +
	\frac{N-1}{N} n_m^{*2}\hat{\pi} (1-\hat{\pi}) \hat{\rho} \Big]}.
\end{equation*}
Internally, the estimate of the binomial proportion is given as $\hat{\pi}= \sum_{h=1}^{H} y_h /
\sum_{h=1}^{H} n_h$ and the estimate for the intra-class correlation $\hat{\rho}$
is calculated following \citet{Lui2000}.
The bootstrap calibration is done using the algorithm given in section \ref{sec::bs_calibration},
with $\boldsymbol{y}_b$ and $\boldsymbol{y}^*_b$ sampled using \texttt{rbbinom()}
which is described in section \ref{sec::sampling}. Please note,
that for the user, the functionallity of \texttt{beta\_bin\_pi()} is exactly the same
as of \texttt{quasi\_bin\_pi()}, meaning that the handling of historical and future
data does not differ from each other. Furthermore, the output of both functions
has the same format.\\
A prediction interval for the number of success in one future cluster of size 50
can be obtained with
\begin{verbatim}
R> set.seed(1234)
R> beta_bin_pi(histdat=qb_dat1, newsize=50, nboot=10000)
\end{verbatim}
\begin{verbatim}
  total hist_prob quant_calib  pred_se lower    upper
1    50      0.13    2.429453 4.395622     0 17.17896
\end{verbatim}
Simultanious prediction intervals for the numbers of success out of three clusters
of size 40, 50 and 60 can be obtained by
\begin{verbatim}
R> set.seed(1234)
R> beta_bin_pi(histdat=qb_dat1, newsize=c(40, 50, 60), nboot=10000)
\end{verbatim}
\begin{verbatim}
  total hist_prob quant_calib  pred_se lower    upper
1    40      0.13    3.405039 3.643114     0 17.60495
2    50      0.13    3.405039 4.395622     0 21.46727
3    60      0.13    3.405039 5.144237     0 25.31633
\end{verbatim}
If a future data set (in this case \texttt{predint::bb\_dat2}) is available, it can be
specified via \texttt{newdat}
\begin{verbatim}
R> set.seed(1234)
R> beta_bin_pi(histdat=qb_dat1, newdat=bb_dat2, nboot=10000)
\end{verbatim}
\begin{verbatim}
  succ fail total hist_prob quant_calib  pred_se lower    upper cover
1   11   29    40      0.13    3.405039 3.643114     0 17.60495  TRUE
2    1   49    50      0.13    3.405039 4.395622     0 21.46727  TRUE
3    3   57    60      0.13    3.405039 5.144237     0 25.31633  TRUE
\end{verbatim}


\subsubsection{Prediction intervals for overdispersed Poisson data}

Bootstrap calibrated prediction intervals for overdispersed Poisson data are implemented
in \texttt{quasi\_pois\_pi()} and are calculated as
\begin{equation}
	[l,u]=\hat{\lambda} \pm \delta \sqrt{\hat{\phi} \hat{\lambda} \Big(1 + \frac{1}{H} \Big)}
\end{equation}
with $\delta$ as the calibrated coefficient used in eq. \ref{eq::bs_calib_pi}.
Please note, that the sampling of bootstrap data $\boldsymbol{y}_b$  and $\boldsymbol{y}^*_b$
in step two and three of the calibration process is done based on \texttt{rqpois()},
which will be described below in section \ref{sec::sampling}.
The data set \texttt{qp\_dat1} contains sampled data that mimics historical
observations (e.g. eggs per hen over two years) obtained from several clusters
(e.g. studies).
\begin{verbatim}
R> qp_dat1
\end{verbatim}
\begin{verbatim}
[1] 46 62 30 59 74 53 32 27 59 47
\end{verbatim}
A prediction interval for one future observation is given by
\begin{verbatim}
R> set.seed(1234)
R> quasi_pois_pi(histdat=data.frame(qp_dat1), m=1, nboot=10000)
\end{verbatim}
\begin{verbatim}
  m hist_mean quant_calib  pred_se    lower    upper
1 1      48.9    2.253848 16.23642 12.30559 85.49441
\end{verbatim}
Please note, that the historical data specified via \texttt{histdat} needs to be a
\texttt{data.frame}. The number of future observations that should be covered by
the prediction interval can be specified by \texttt{m}.
A prediction interval for $M=3$ future observations can be obtained by

\begin{verbatim}
R> set.seed(1234)
R> quasi_pois_pi(histdat=data.frame(qp_dat1), m=3, nboot=10000)
\end{verbatim}
\begin{verbatim}
  m hist_mean quant_calib  pred_se lower    upper
1 3      48.9    3.092852 16.23642     0 99.11683
\end{verbatim}

If the future data is already available (here \texttt{qp\_dat2}), it can be specified
via \texttt{newdat}
\begin{verbatim}
R> qp_dat2
\end{verbatim}
\begin{verbatim}
[1] 44 74 36
\end{verbatim}
\begin{verbatim}
R> set.seed(1234)
R> quasi_pois_pi(histdat=data.frame(qp_dat1),
                 newdat=data.frame(qp_dat2),
                 nboot=10000)
\end{verbatim}
\begin{verbatim}
  qp_dat2 hist_mean quant_calib  pred_se lower    upper cover
1      44      48.9    3.092852 16.23642     0 99.11683  TRUE
2      74      48.9    3.092852 16.23642     0 99.11683  TRUE
3      36      48.9    3.092852 16.23642     0 99.11683  TRUE
\end{verbatim}


\subsection{Functions for data sampling and bootstrapping} \label{sec::sampling}

Since, all prediction intervals implemented in \textbf{predint} are based on bootstrap
calibration, functions for the sampling of new observations from the models described above
are necessary. An overview about these functions is given in table \ref{table:sampling}.


\begin{table}[h!]
\centering
\caption{Sampling functions implemented in \textbf{predint}}
\label{table:sampling}
\begin{tabular}{l|l}
\hline
 \textbf{Function name} & \textbf{Functionality}  \\ \hline
 \texttt{lmer\_bs()} & Bootstrapping from random effects models \\ \hline
 \texttt{rbbinom()} &  Sampling of beta-binomial data  \\ \hline
 \texttt{rqbinom()} &  Sampling of quasi-binomial data \\ \hline
 \texttt{rqpois()} &  Sampling of quasi-Poisson data    \\ \hline
\end{tabular}
\end{table}


\subsubsection{Bootstrapping from random effects models} \label{sec::lmer_bs}

In principle, bootstrapping from linear random effects models fit with \texttt{lme4::lmer()}
can be done with \texttt{lme4::bootMer()}. Anyhow, the bootstrap samples obtained
with \texttt{lme4::bootMer()} are bound to have the same experimental structure (same
numbers of observations per random factor) as the original data set the model was
fit to. \\
As already stated in section \ref{sec::ran_eff_mod}, a simultainious prediction intervall
for $M > 1$ future observations depends on the numbers of observations per
random factor in the future data set. Hence, the variance-covariance matrices
$\boldsymbol{V}$ and $\boldsymbol{V^*}$ for the historical and the future observations
usually differ from each other. \\
A bootstrap function, that is able to sample new data sets based on the estimated
mean and variance components drawn from a random
effects model fit with \texttt{lme4::lmer()}, in which the bootstraped data does
not have to be of the same structure as the initial data, is provided via \texttt{lmer\_bs()}
and is based on the sampling algorithm described in section \ref{sec::PI_ranef}.\\
\texttt{lmer\_bs()} depends on the following arguments: \texttt{model, newdat, futmat\_list}
and \texttt{nboot}.\\
\texttt{model} defines the random effects model fit with \texttt{lme4::lmer()}. Please
note, that \texttt{lmer\_bs()} only works for models in which random effects are
specified as \texttt{(1 | random effect)}.
\texttt{nboot} defines the number of bootstrap samples $B$.
If \texttt{newdat} is defined, the design matrices $\boldsymbol{Z}^*_c$ are computed
using \texttt{lme4::lFormula}. But, as described before, \texttt{lme4::lFormula} requires
at least to observations per random factor. If this is not the case, a list containing the
design matrices can be supplied via \texttt{futmat\_list}.
Based on the fitted model

\begin{verbatim}
R> fit <- lmer(y_ijk~(1|a)+(1|b)+(1|a:b), c2_dat1)
\end{verbatim}

100 bootstrap samples that have the same experimental structure as \texttt{c2\_dat3}
can be sampled with

\begin{verbatim}
R> lmer_bs(model=fit, newdat=c2_dat3, nboot=100)
\end{verbatim}

Alternatively new data can be sampled based on a list that contains the design
matrices $\boldsymbol{Z}^*_c$ that can be specified via \texttt{futmatlist}

\begin{verbatim}
R> lmer_bs(model=fit, futmat_list=fml, nboot=100)
\end{verbatim}
with \texttt{fml} defined above in section \ref{sec::PI_ranef}.


\subsubsection{Sampling of beta-binomial data}

If the data is assumed to be beta-binomial distributed, such that
\begin{gather*}
	\pi_i \sim beta(a, b) \\
	y_i \sim bin(\pi_i, n_i) \\
	var(y_i)^{BB}= n_i \pi (1-\pi) [1+(n_i-1) \rho]
\end{gather*}
with $i=1, \ldots, I$ clusters of size $n_i$ and intra-class correlation coefficinent
\begin{equation*}
	\rho=\frac{1}{1+a+b},
\end{equation*}
it can be sampled using the following mechanism:\\
Based on given values for $\pi$ and $\rho$, the parameters of the beta-distribution
$a$ and $b$ can be calculated as
\begin{gather*}
	a+b = \frac{1 - \rho}{\rho} \\
	a=\pi (a+b) \\
	b=(a+b)-a.
\end{gather*}
with $\pi=E(\pi_i)=a/(a+b)$. Then, the binomial
proportions for each cluster are sampled from the beta distribution
\begin{equation*}
	\pi_i \sim Beta(a, b)
\end{equation*}
and the numbers of successes for each cluster are sampled to be
\begin{equation*}
	y_i \sim Bin(n_i, \pi_i)
\end{equation*}
for a given cluster size $n_i>1$.
Please note, that this sampling mechanism only works if $\rho$ is bigger than zero
but smaller than one.\\
This approach is implemented in \texttt{rbbinom()} in which \texttt{n} refferes to
the number of clusters $I$, \texttt{size} refferes to the cluster
size $n_i$, \texttt{prob} referes to the expected binomial proportion $\pi$ and
\texttt{rho} to the intra class correlation coefficient $\rho$.\\
A data set with ten clusters, each comprized of 50 experimental units, an expected
success probability of 0.1 and an intra class correlation of 0.06 can be sampled
as
\begin{verbatim}
R> rbbinom(n=10, size=50, prob=0.1, rho=0.06)
\end{verbatim}


\subsubsection{Sampling of quasi-binomial data} \label{sec::rqbinom}

Quasi-binomial data sampling is based on the assumption that the binomial
variance is inflated by a dispersion paramter that is constant for all $i=1, \ldots I$
clusters
\begin{equation*}
	var(y_i)= \phi n_i \pi (1-\pi).
\end{equation*}
This type of data can be sampled from the beta-binomial distribution using the
following mechanism:\\
For a given cluster size $n_i > 1$ and a given dispersion parameter $\phi$,
the sum of the parameters of the beta-distribution differs between the $i=1, \ldots I$
clusters and is given by
\begin{equation}
	(a+b)_i=\frac{\phi-n_i}{1-\phi} \label{eq::asum}.
\end{equation}
Subsequently, $a_i$ and $b_i$ can be calculated individually for each cluster,
based on a predifined value between 0 and 1 for $\pi$
\begin{gather*}
	a_i=\pi (a+b)_i \\
	b_i=(a+b)_i-a_i.
\end{gather*}
Then, the binomial proportions for each cluster are sampled
from individual beta distributions
\begin{equation*}
	\pi_i \sim Beta(a_i, b_i)
\end{equation*}
and the numbers of succes for each cluster are sampled to be
\begin{equation*}
	y_i \sim Bin(n_i, \pi_i).
\end{equation*}
Please note, that this sampling mechanism works only, if $\phi > 1$ and
$\phi < n_i$. Both, a dispersion parameter of $\phi < 1$ as well as $\phi > n_i$
result in negative $(a+b)_i$ in eq. \ref{eq::asum} and hence in negative
values for $a_i$ and $b_i$ as well. Furthermore, $\phi=1$ and $\phi = n_i$
result in $(a+b)_i=0$. Anyhow, the beta-distribution is only defined if their
parameters are positive numbers greater than zero.\\
An implementation of this sampling process is provided via \texttt{rqbinom()} wich
depends on the arguments \texttt{n, size, prob} and \texttt{phi}. Similar to \texttt{rbbinom()},
\texttt{n} refers to the number of clusters $I$, \texttt{size} to the size of the clusters $n_i$
and \texttt{prob} to the expected binomial proportion $\pi_i$. \texttt{phi} defines the dispersion
parameter $\phi$.\\
A data set with ten clusters, each comprized of 50 experimental units, an expected
success probability of 0.1 and a dispersion parameter of three can be sampled as
\begin{verbatim}
R> rqbinom(n=10, size=50, prob=0.1, phi=3)
\end{verbatim}


\subsubsection{Sampling of quasi-Poisson data}

The sampling of quasi-Poisson data is based on the assumption, that the dispersion
parameter constantly inflates the variance of the observations obtained in
$i=1, \ldots, I$ clusters, such that
\begin{gather*}
	var(y_{i}) = \lambda (1 + \lambda \kappa) =\phi \lambda
\end{gather*}
as described above (see eq. \ref{eq::var_QB_NB}). Hence overdispersed Poisson data
with constant overdispersion can be sampled from the negative-binomial distribution
as follows:
Define $\kappa$ as
\begin{equation*}
	\kappa = \frac{\phi - 1}{\lambda}
\end{equation*}
for given values of $\phi > 1$ and $\lambda > 0$. Then calculate
\begin{gather*}
	a=\frac{1}{\kappa} \\
	b=\frac{1}{\kappa \lambda}
\end{gather*}
and sample the poisson means for each cluster from the gamma distribution, such that
\begin{equation*}
	\lambda_i \sim Gamma(a,b).
\end{equation*}
Subsequently, the observations are sampled from the Poisson distribution
\begin{equation*}
	y_i \sim Pois(\lambda_i).
\end{equation*}
This sampling process is implemented in \texttt{rqpois()} which depends on the arguments
\texttt{n}, \texttt{lambda} and \texttt{phi}. Similar to \texttt{rbbinom()} and \texttt{rqbinom()},
\texttt{n} referes to the number of clusters $I$. The expected value for the observations
$E(y_i) = \lambda$ is defined by \texttt{lambda} and the dispersion parameter $\phi$ by \texttt{phi}.
Hence, a data set with ten clusters, an expected value of fife and dispersion parameter
of three can be sampled as

\begin{verbatim}
R> rqpois(n=10, lambda=5, phi=3)
\end{verbatim}


\section{Summary} \label{sec:summary}

The \textbf{predint} package is the first R-package available from CRAN that provides
prediction intervals for $M \geq 1$ future observations based on random effect models,
overdispersed binomial data or based on overdispersed poisson data. Although
the implemented methodology evolved from applications in the context of toxicology
and pre-clinical statistics \citep{Menssen+Schaarschmidt:2019, Menssen+Schaarschmidt:2021},
it might be applicable in a broad range of other research fields. \\

\subsection{Interpretation of the implemented prediction intervals}

A prediction interval for $M=1$ future observation can be interpreted as a
\textit{pointwise} prediction interval. This kind of interval should cover one
future observation in $(1-\alpha)$\% of the cases. If such an interval is applied based on an
univariate distribution, it directly approximates the central $(1-\alpha)$\% of
this distribution. This is because both, the historical as well as the future
observation(s) are believed to be independend realisations of exactly the same
distribution. Hence, in the long run, its borders $L$ and $U$ converge to the
$\alpha / 2$ and the $1-\alpha / 2$ quantiles of the undelying distribution
(see \citealt{Francq_et_al:2019} Fig. 1).\\
In this special case, such an prediction interval can be interpreted
as a $\beta$-content tolerance interval that covers the central $(1-\alpha)$\%
of the distribution. A univariate prediction interval for $M=1$ future observation,
that is based on one normal distributed sample, is revieved in \citet{Hahn+Meeker+Escobar:2017}
and implemented in the \textbf{BivRegBLS} package of \citet{Francq+Berger+Boachie:2020}
in order to be displayed in a Bland-Altman plot.\\
Anyhow, if the underlying data is comprised of several clusters (e.g. due to
repeated measurements), the sample is not comprised of independend observations
anymore. These dependencies can be taken into account, if the data is modeled
by a random (or mixed) effects model (see section \ref{sec::ran_eff_mod}). This
approach leads to the assumption that the historical and the future data usually
do not follow exectly the same distribution, although they descent from the same
data generating process (see eq. \ref{eq::Y_mvn} and \ref{eq::Y_star_mvn}).
In this case, one has to be extremely carefull, if the prediction interval
for $M=1$ future observation is aimed be interpreted as a $\beta$-content tolerance
interval, since the distribution of a future sample changes with its experimental
design.\\
\textit{Simultanious} prediction intervals should cover all of $M>1$ future observations
and hence are applicable to a broad field of applications in which, at the moment,
the application of tolerance intervals seems to be favoured (e.g. in toxicology
or antidrug-antibody detection). Furthermore, the problem that in random (or mixed)
effects models the historical and the future observations usually follow different
distributions, should also influence the calculation and
interpretation of $\beta \gamma$-tolerance intervals (which should cover the central
$100\gamma$\% of the underying distribution with coverage probability $\beta$). Hence,
further theoretical work will follow on that topic.


\subsection{The future of predint}

It is planned, that the future research on prediction intervals (as well as on other intervals)
will be included in \textbf{predint}, if it fits into the initial scope of this
package (the use of historical control data for the validation of actual observations).\\
Hence, the implemented methodology for random effects models will be extended
to be also applicable in the context of models with mixed effects. This kind of models
are of use, if a data set contains several factors of interest, such as the strain
or the sex of rats as well as several random factors, representing the experimental
design. At the moment, it seems to be common to split such a data set
according to the factors of interest (e.g. female rats of a given strain) and
assume that these subsets are samples of independent observations
\citep{Igl_et_al:2013, Menssen+Schaarschmidt:2019, Elmore+Peddada:2009}.
As stated above, an alternative approach that is based on the complete data set
is the application of mixed effects models. But, prediction intervals that are based on
such models are not available in an R package so far. Therefore, it is planned
to fill this gab. Furthermore, it is planned to implement tolerance intervals
for both, random and mixed effects models.


%


\section*{Acknowledgments}
I have to thank Frank Schaarschmidt for his time he spend discussing 
the ideas behind the methodology that finaly eveolved to become the \textbf{predint}
package.


\bibliography{predint_paper}


\end{document}